\documentclass[10pt, conference, compsocconf]{IEEEtran}
\ifCLASSINFOpdf
\else
\fi
\usepackage{cite}
\usepackage{amsmath,amssymb,amsfonts}
\usepackage{graphicx}
\usepackage{cite}
\usepackage{tabularx}
\usepackage{booktabs}
\usepackage{amsmath,amssymb,amsfonts}
\usepackage{algorithmic}
\usepackage{textcomp}
\usepackage{listings,xcolor}
\usepackage{url}
\def\BibTeX{{\rm B\kern-.05em{\sc i\kern-.025em b}\kern-.08em
    T\kern-.1667em\lower.7ex\hbox{E}\kern-.125emX}}
\begin{document}

\title{Scoring Popularity in GitHub}

\author{\IEEEauthorblockN{Abduljaleel Al-Rubaye}
\IEEEauthorblockA{\textit{Department of Computer Science} \\
\textit{University of Central Florida}\\
Orlando, FL US \\
aalrubaye@knights.ucf.edu}
\and
\IEEEauthorblockN{Gita Sukthankar}
\IEEEauthorblockA{\textit{Department of Computer Science} \\
\textit{University of Central Florida}\\
Orlando, FL US \\
gitars@eecs.ucf.edu}

}

\maketitle              
\begin{abstract}
Popularity and engagement are the currencies of social media platforms, serving as powerful reinforcement mechanisms to keep users online.   Social coding platforms such as GitHub serve a dual purpose: they are practical tools that facilitate asynchronous, distributed collaborations between software developers while also supporting passive social media style interactions.  There are several mechanisms for ``liking'' content on GitHub: 1) forking repositories to copy their content  2) watching repositories to be notified of updates and 3) starring to express approval. This paper presents a study of popularity in GitHub and examines the relationship between these three quantitative measures of popularity.  We introduce a weight-based popularity score ($WTPS$) that is extracted from the history line of other popularity indicators.
\end{abstract}
\IEEEpeerreviewmaketitle

\section{Introduction}
 With more than 31 million users and 2.1 million organizations, GitHub is one of the largest social coding platforms~\cite{githubOctoverse}. Users store their files in repositories which can be shared with others; GitHub currently hosts more than 96 million repositories. User interactions engender events that help the repositories perform version control. For instance, the \textit{Create} event is used to create a repository, the \textit{Fork} event is triggered when a repository is cloned by a user, and the \textit{Push} event occurs when a user pushes files to a repository~\cite{github}.  Beyond simple version control, GitHub also supports social engagement between users. Users can express approval of content by starring it or can opt to be notified of content changes by watching repositories.

 These measures implicitly serve as a recommendation of the ideas, methods, innovation, and reusability of the content. Being an author of a repository that is frequently forked is similarly to being highly cited. GitHub has its own gh-impact score that is comparable to h-index. Three  different measurements of popularity are provided: \textit{forks}, \textit{stars}, and \textit{watchers}.  The \textit{fork} count shows the number of people who were sufficiently interested to clone or copy a repository on their local resource, while \textit{stars} indicate users who have agreed to receive notifications about all repository activities. \textit{Watchers}, similarly, represent the number of users who bookmarked a repository. This paper reports on patterns and trends of popularity and social engagement in GitHub. Popularity measures make it easier for users to rapidly find highly recommended content. In the quest to achieve higher ratings, repository owners may be motivated to improve the quality of the hosted work to attract more people.  The dark side of popularity ratings is that they may suppress users' creativity, leading them to remove content that is unpopular. For this reason, Instagram actually obscures the popularity of content to reduce users' anxiety.  The next section discusses related work on popularity in social coding platforms.


\section{Related Work}
Most studies conducted by mining data from social coding platforms have relied on a single popularity metric, such as stars, to quantify repository popularity. Xavier et al.~\cite{xavier2014understanding} examined the number of developer followers and noted that commit activity is definitely a factor contributing to higher developer popularity.  Popularity in other domains, such as authoring blogs or activity on other social networks, can also contribute to GitHub developer popularity.

Borges et al. have conducted several studies on GitHub popularity, analyzing the main factors that can impact starring activity~\cite{borges2016understanding}.  For instance, they looked at the relationship between repository age, number  of  commits, contributors, forks and starring activity.   Moreover they studied patterns of popularity growth;   based on their work on growth patterns, they proposed a multiple linear regression technique for predicting the number of future stars a repository will receive~\cite{borges2016predicting}. Al Rubaye and Sukthankar~\cite{alrubaye2018} proposed a model to predict the diffusion of innovation in GitHub. During the model creation process, they used forks as an indicator of repository popularity. 

In addition to these studies there are several online tools that rank repositories by different popularity measurements. For instance, GitHub itself ranks the rock-star repositories in a descending order by their daily stars only \cite{githubranking}. \textit{Git Awards} \cite{gitawards} is a web-based application that gives a repository's ranking on GitHub by language or by geolocation based on the number of stars. \textit{Git Most Wanted} \cite{gitmostwanted} is another web-based application that ranks the repositories based on stars or fork counts.

\section{Method}
\subsection{Data Set}
To prepare our data set, we utilized the online archive of public repositories provided by the project GHTorrent \cite{ghtorrent} which extracts all the data using GitHub's REST API. For this work 36,000 public repositories were randomly captured on October 30th, 2018. 

\begin{figure}[htbp]
\centerline{\includegraphics[width=0.6\columnwidth]{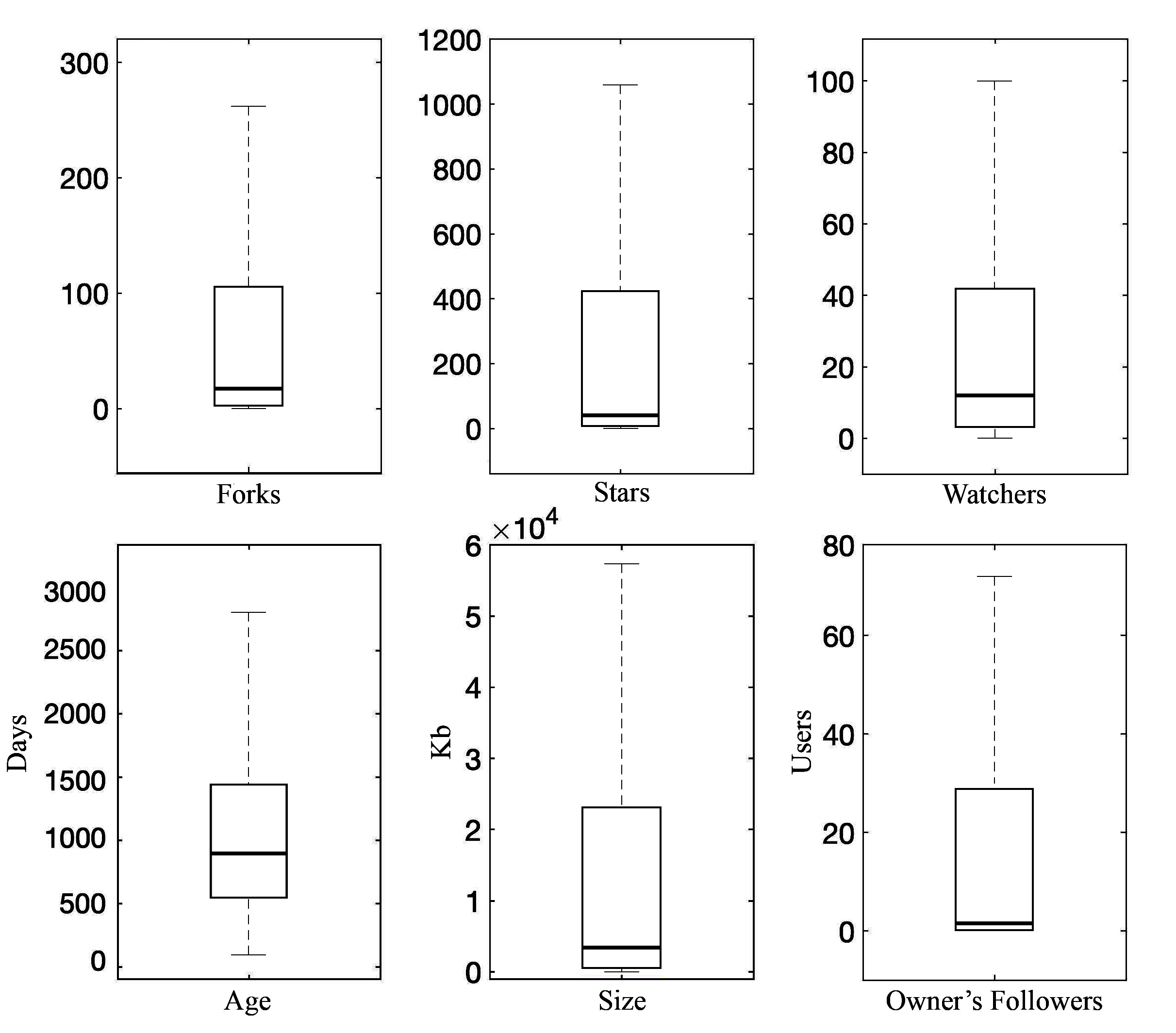}}
\caption{Distribution of repository features: forks, stars, watchers, age of repositories, size, and number of repository owner followers.}
\label{boxplots}
\end{figure}

Each record contains a repository object including all the endpoints of the related events, commits, contributors, etc. The data objects also provide repository features and information such as the creation date, the primary programming language, size of the repository, fork count, watcher count, star count, and repository owner information. Figure  \ref{boxplots} presents the distribution of the key properties of our data set. To extract the timeline of occurrences for popularity-related events, we used the property \textit{timestamp} that was provided by GitHub API to retrieve the exact time points that a fork or star event occurred. 


\subsection{Popularity Growth}

\begin{figure*}[ht]
	\begin{center}
		\includegraphics[width=0.5\textwidth]{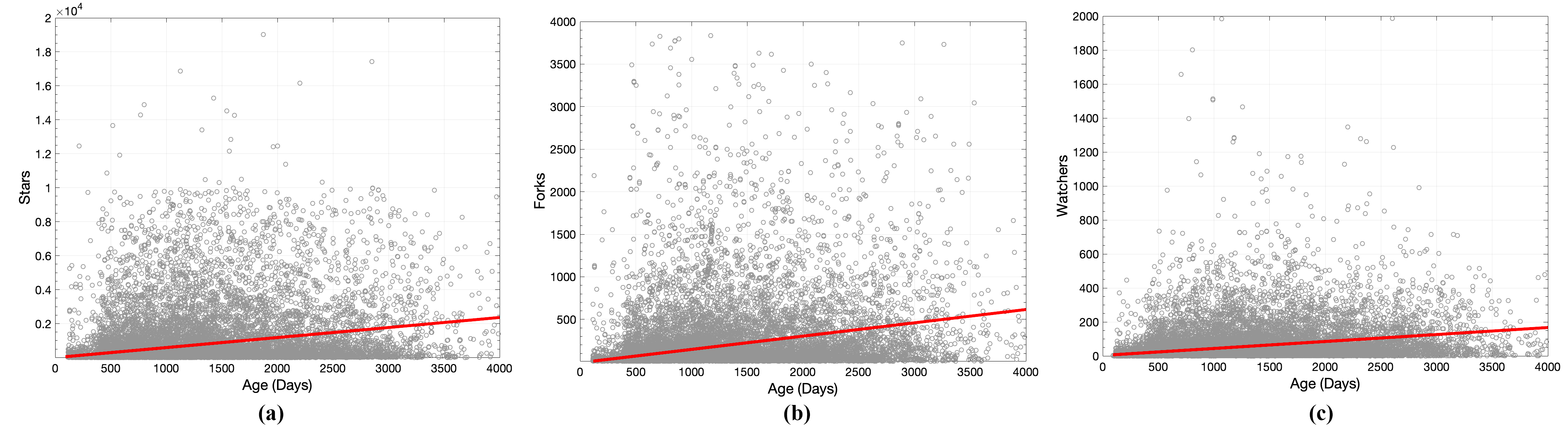}
	\caption{Correlation of age and popularity measurements. (a) shows the age-star correlation (coefficient=0.3215). (b) shows the age-fork correlation (coefficient=0.2923) and (c) shows age-watcher correlation (coefficient=0.3162). \cite{lineregression}}
	\label{age_relations}
	\end{center}
\end{figure*}

The concept of popularity cannot be separated from time.  Figure \ref{age_relations} depicts the correlation between repository age (in days) and  popularity indicators; obviously over time, we see an increase in total number of people who fork, star or watch a repository.

\begin{figure}[htbp]
\centerline{\includegraphics[width=0.7\columnwidth]{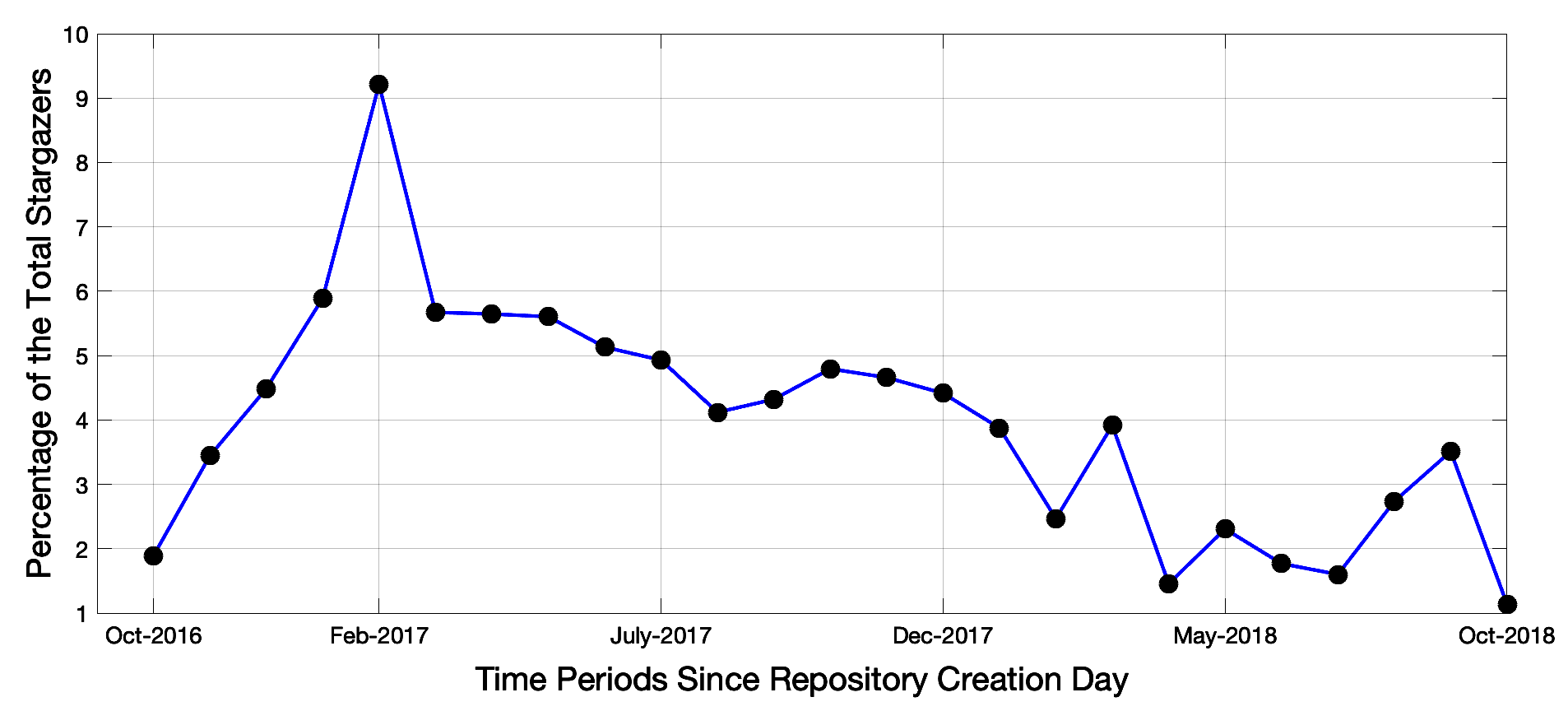}}
\caption{An example of repository star growth patterns}
\label{timeline}
\end{figure}

Usually repositories experience a spike of popularity growth. For instance, the repositories that are created by big tech companies and well-known organizations may attract  developers in a single burst. Figure \ref{timeline} demonstrates an example of such a repository, which received more than 9\% of its total stargazers count on February 2017 and was created on October 2016.

Repositories can be grouped in three categories based on their popularity growth over time. The first group includes the repositories that have only attracted a small number of developers. They fail to raise their popularity to a noticeable point and remain at the same level for a long time. Consequently this type of repository experiences litttle growth. Some repositories  gain popularity via various means but then lose it when some of GitHub users \textit{unwatch} or \textit{unstar} them. The last group consists of  repositories that keep gaining popularity, where their overall growth rate is positive over the majority of their life time. These repositories may be considered as resources of innovation for developers on GitHub~\cite{borges2016understanding}.


\subsection{Calculating Popularity Score}
Popularity in GitHub can be viewed as a measurement of how attracted users are to the repository content.  This aura of attractiveness may not be completely captured by the number of watchers, stargazers or forkees.  We propose the use of a weighted popularity score to more accurately express different aspects of popularity.

We illustrate the benefits of other scoring techniques with an example.
Imagine there is a repository (A) that is forked by 1000 users and starred by 20 and repository (B) that is not forked at all yet but has been starred 1020 times. Considering star count as the popularity indicator, repository (B) is more popular, but if only fork counts are measured, the opposite is true---repository (A) is more popular. We believe that it is useful to model repository growth patterns, examining the timeline of increases in forks and stars. Some may be more stable in gaining popularity and others may fluctuate over time, gaining thousands of stars/forks in a month and only a few stars/forks on the next month. The benefit of a weighted score is its ability to model these factors.

\textit{\textbf{Popularity Weight}}:
Popularity quantifies interactions between a community's components; it is almost a meaningless concept when considering an element in isolation.
A set of repositories can be considered a community, in which the popularity of a repository should be considered relative to other repositories. In other words, if a repository receives more stars than other repositories during a time period, this may be more significant than achieving a higher absolute count of stars.

The idea of time-based popularity weights emerges from the fact that during some time periods we see thousands of developers forking, starring, or watching different repositories whereas during other periods the total number of popularity related events is much lower. Hence we treat these periods as having different popularity weights. Consequently, gaining forks, stars or watchers at certain times may be more valuable than at others. 
We introduce the \textit{Weighted Total Popularity Score ($WTPS$)} to quantify the popularity of a repository that is part of a community; $WTPS$ weights popularity with respect to time. Initially the time interval $t$ is defined to be equal to one month. Therefore, in order to find the popularity indicators' weights we take all the forks and stars that were captured on each month, then we calculate the weights against the total gained forks and stars of the repositories of our data set. Weights of forks ($W_{F_t}$) and stars ($W_{S_t}$) for each time interval $t$ are calculated as follows:

\begin{equation}
W_{F_t}=\frac{\sum_{i=0}^{n}Forks[R_i]_t}{\sum_{i=0}^{n}Forks[R_i]}
\label{CalculatingForkWeights}
\end{equation}
\begin{equation}
W_{S_t}=\frac{\sum_{i=0}^{n}Stars[R_i]_t}{\sum_{i=0}^{n}Stars[R_i]}
\label{CalculatingStarsWeights}
\end{equation}

such that $\sum_{i=0}^{n}W_{F_t}=1$ and $\sum_{i=0}^{n}W_{S_t}=1$, where $n$ is the total number of the repositories. Note that to compute the popularity score of an individual repository without considering the fact that it is part of a larger community of repositories on GitHub is equivalent to setting $W_{F_t}$ and $W_{S_t}$ to be equal to 1. Table \ref{tab:exampleTable1} shows a sample data set of repositories and their captured forks and stars at different time intervals, along with their calculated $W_{F_t}$ and $W_{S_t}$ values. By calculating the fork and star weights of each time interval, we are able to calculate the weighted total popularity score $WTPS$ for each repository $R$.
\smallskip
\smallskip
\smallskip
\smallskip
\begin{table*}[]
\centering
\caption{A sample data set of four repositories and their collected forks and stars at 5 time points [$t_1$... $t_5$].  Weights $W_F$ and $W_S$ were calculated using Equations \ref{CalculatingForkWeights} and \ref{CalculatingStarsWeights}}.

\label{tab:exampleTable1}
\begin{tabular}{l|lllll|l|lllll|l|}
\cline{2-13}
\multicolumn{1}{l|}{}& $F_{t_1}$    & $F_{t_2}$    & $F_{t_3}$    & $F_{t_4}$    & $F_{t_5}$    & $F_{total}$ & $S_{t_1}$     & $S_{t_2}$     & $S_{t_3}$     & $S_{t_4}$     & $S_{t_5}$     & $S_{total}$ \\ \hline
\multicolumn{1}{|c|}{$R_1$} & 15     & 20     & 3      & 9      & 7      & 54     & 6      & 12     & 2      & 15     & 10     & 45     \\
\multicolumn{1}{|c|}{$R_2$} & 10     & 10     & 5      & 3      & 30     & 58     & 6      & 5      & 6      & 5      & 8      & 30     \\
\multicolumn{1}{|c|}{$R_3$} & 5      & 8      & 4      & 10     & 15     & 42     & 1      & 3      & 4      & 22     & 20     & 50     \\
\multicolumn{1}{|c|}{$R_4$} & 14     & 12     & 10     & 5      & 5      & 46     & 12     & 14     & 10     & 13     & 6      & 55     \\ \hline
\multicolumn{1}{|c|}{Weight} & $\frac{44}{200}$ & $\frac{50}{200}$ & $\frac{22}{200}$ & $\frac{27}{200}$ & $\frac{57}{200}$ & $1$      & $\frac{25}{180}$ & $\frac{34}{180}$ & $\frac{22}{180}$ & $\frac{55}{180}$ & $\frac{44}{180}$ & $1$ \\ \hline
\end{tabular}
\end{table*}

\begin{table}[ht]
\begin{center}
	\caption{Calculating the Weighted Total Popularity Score $(WTPS)$ for the repositories detailed in Table \ref{tab:exampleTable1} based on Equation \ref{WTPTOverall}}
	\label{tab:exampleTable2}
	\begin{tabular}{l|llll}
		\toprule
            $WTPS$& $R_1$    & $R_2$    & $R_3$    & $R_4$ \\ \hline
            $t_1$       & 4.08     & 2.98     & 1.23      & 4.64    \\
            $t_2$         & 7.25     & 3.44     & 2.56      & 5.63  \\
            $t_3$         & 0.57      & 1.28      & 0.92      & 2.32\\
            $t_4$         & 5.71     & 1.9     & 7.93     & 4.57\\
            $t_5$         & 4.43     & 10.5     & 9.15     & 2.89\\ \hline
            $total$         & 22.04     & 20.1     & 21.79     & 20.05\\
		\bottomrule
	\end{tabular}
\end{center}
\end{table}

\textit{\textbf{Weighted Total Popularity Score ($WTPS$)}}:
The equation below shows how to calculate the $WTPS$ of a repository $R$ at time $t$: 
\begin{equation}
WTPS[R_i]_t = (W_{F_t} Forks[R_i]_t) + (W_{S_t}Stars[R_i]_t)
\label{WTPt}
\end{equation}

Hence, by summing up all the weighted popularity scores of a repository for all the time intervals we obtain the overall $WTPS$ as follows:
\begin{equation}
WTPS[R_i]_{overall} = \sum_{t=0}^{m}WTPS[R_i]_t
\label{WTPTOverall}
\end{equation}
where $m$ is the number of the time intervals. Table \ref{tab:exampleTable2} shows the $WTPS$ of each of the repositories of Table \ref{tab:exampleTable1} and at each time point $t1, t2, ...$ as well as the $WTPS_{Overall}$ where both forks and stars are utilized to find the overall popularity score.  The key intuition is that $WTPS$ considers popularity growth relative to the total community. 

\section{Results}
This section presents a comparison of the use of $WTPS$ as an indicator of repository popularity vs.\ the standard measurements.

\subsection{Ranking Popularity}
The previous section described an example set of repositories with different fork/star ratios across several time points. Using our proposed approach of extracting their popularity we calculated the popularity score, $WTPS$ (Table \ref{tab:exampleTable1} and Table \ref{tab:exampleTable2}). 

\begin{figure}[htbp]
	\begin{center}
		\includegraphics[width=0.4\columnwidth]{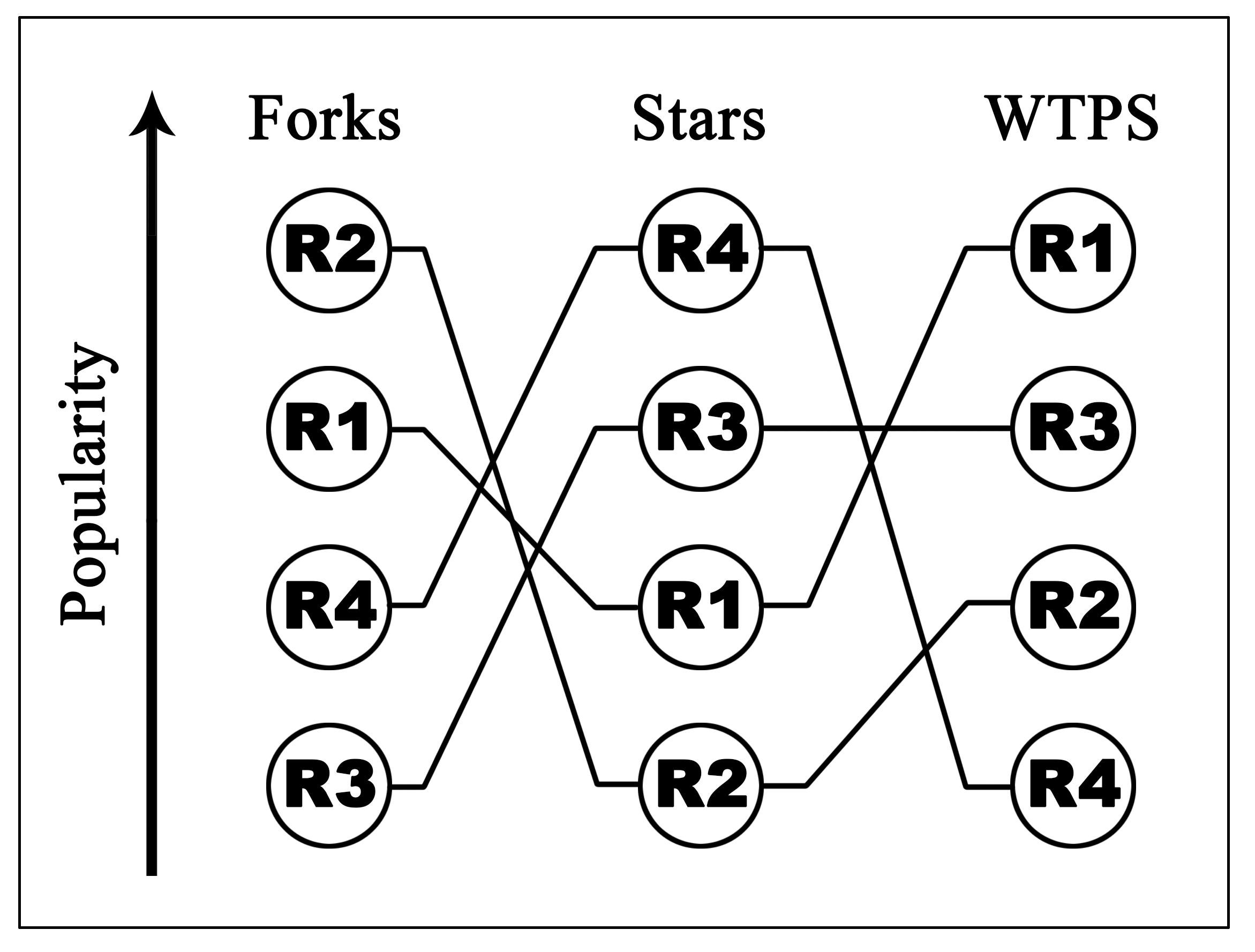}
	\caption{The rank of repositories of Table \ref{tab:exampleTable1} based on forks, stars, and $WTPS$. }
	\label{example_ranks}
	\end{center}
\end{figure}

\begin{figure*}[ht]
	\begin{center}
		\includegraphics[width=0.9\textwidth]{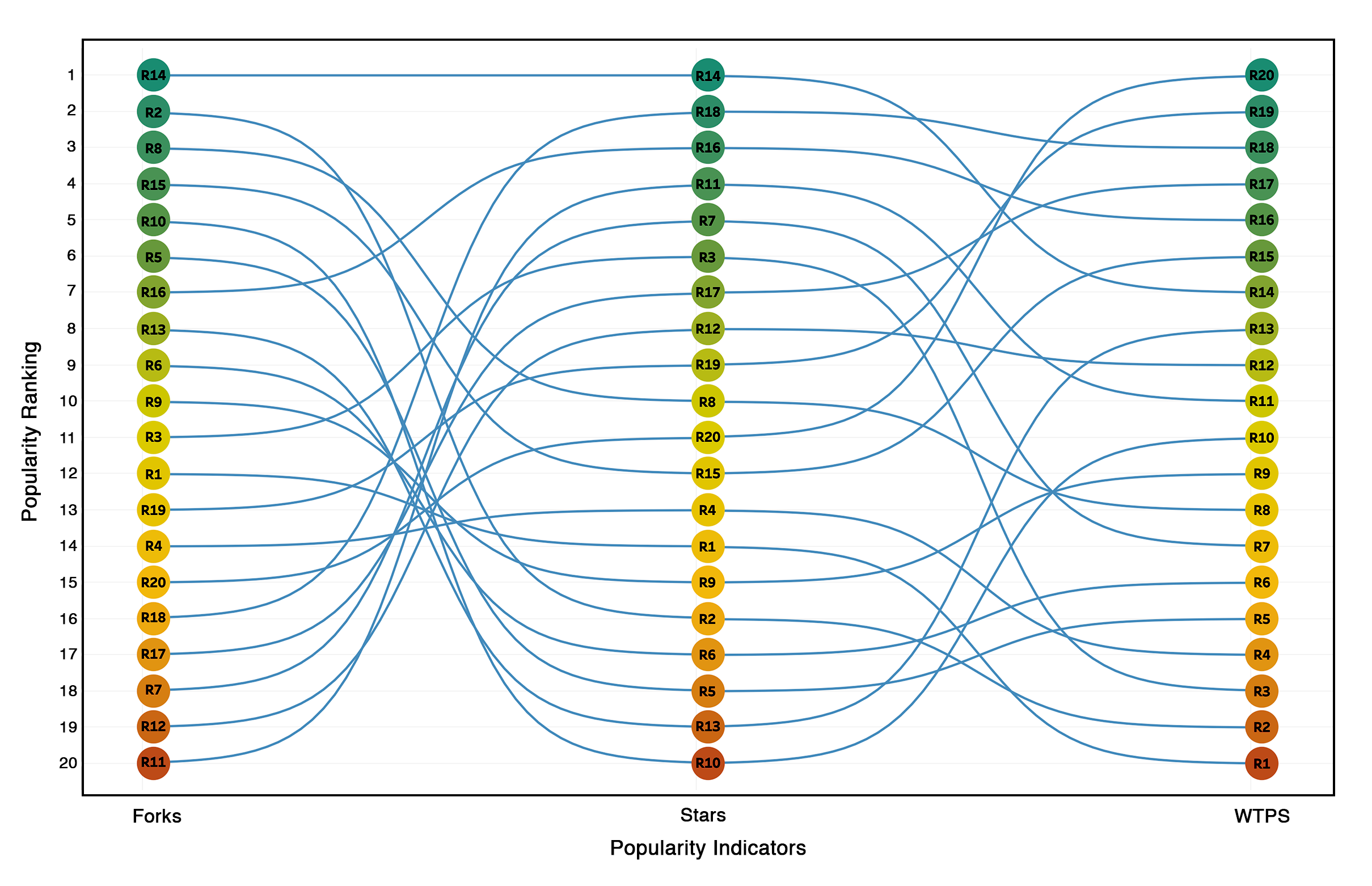}
	\caption{A sample of 20 repositories' popularity ranking over using different popularity measures: forks, stars and $WTPS$ (t=1 month).}
	\label{ranking}
	\end{center}
\end{figure*}

By conducting a simple comparison between a repository's forks and stars against its $WTPS$ we see that its popularity level changes. When the number of accumulated forks or stars of a repository is used as the only popularity indicator, repositories may be ranked either lower or higher than when their popularity score is based on other measurements. A repository that is popular compared to other repositories based on the number of gained forks may have a different rank when star count is used as the popularity indicator. 

Figure \ref{ranking} depicts repository rank changes based on the popularity indicator. For instance, even though $R_{14}$ has the highest ranking among other repositories in fork counts as well as the star counts, it does not have the highest $WTPS$. Since $WTPS$ reweights forks and stars based on their relative standing at different time periods, this results in a lower popularity score for $R_{14}$. Meaning that at some points $R_{14}$ gained forks and/or stars at a time when the popularity weights were relatively lower.  Below we utilize a linear regression approach \cite{lineregression} to analyze the correlation between $WTPS$ and other repository properties more in detail.

\subsection{$WTPS$ vs.\ other popularity indicators}
Clearly the proposed popularity score, $WTPS$, is closely related to forks and stars; however these popularity measurements will be weighted differently over time.  For instance a repository can gain a large number of stars relative to other repositories over a month which will lead to a greater total popularity score. 

\begin{figure}[htbp]
	\begin{center}
		\includegraphics[width=0.8\columnwidth]{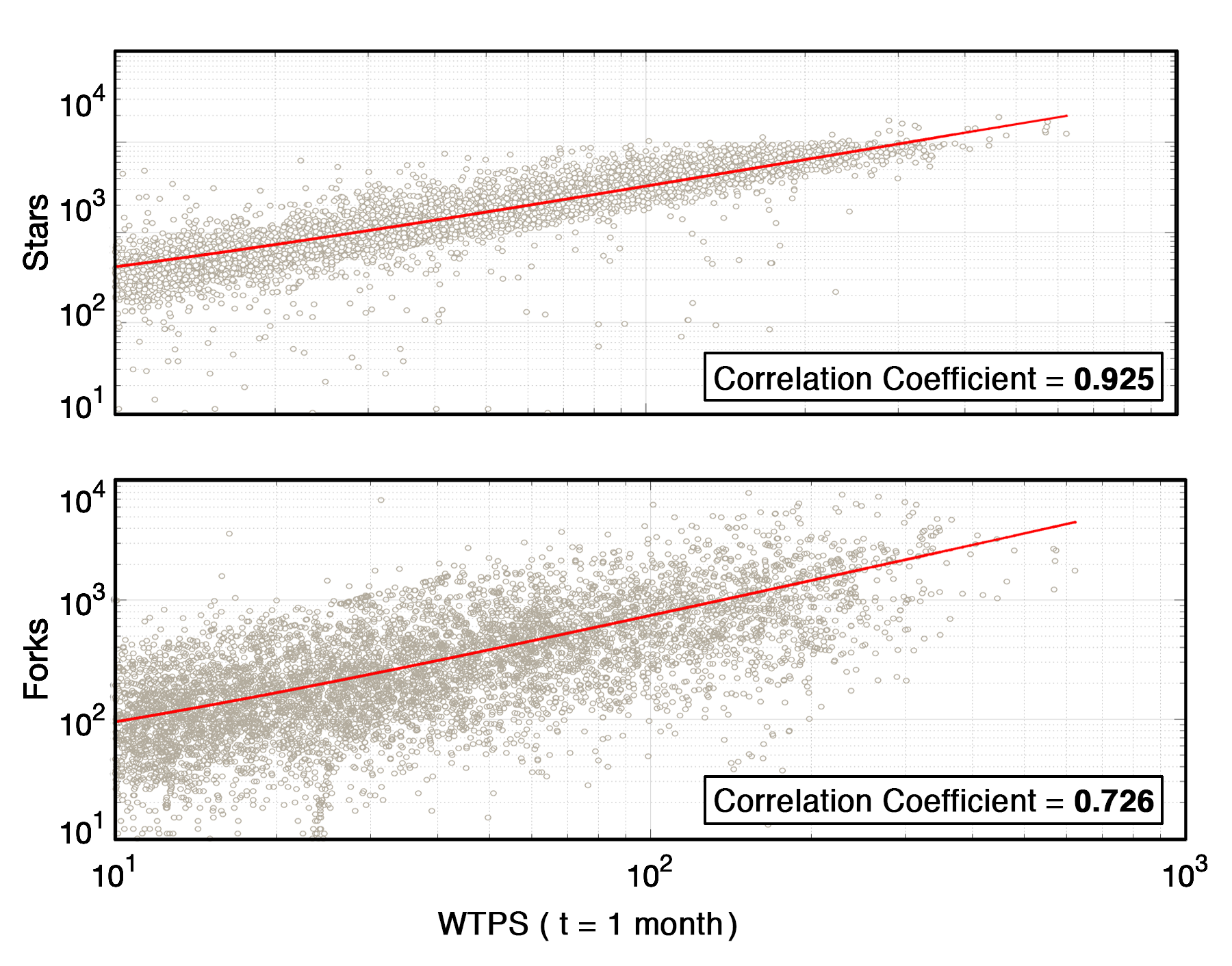}
	\caption{WTPS-Forks and $WTPS$-Stars correlation. }
	\label{mps_forks_stars}
	\end{center}
\end{figure}

To see the overall effect of fork count vs.\ star count on the value of $WTPS$, we compare their correlation to the total $WTPS$ score (Figure \ref{mps_forks_stars}).  It is clearly noticeable that the increase in forks or stars leads to a greater $WTPS$ value in general. 
Even though both indicators are strongly related to the value of $WTPS$, star counts are more correlated with the popularity score values.  For our data set the star count correlation to $WTPS$ is 0.925 which explains why repositories with more stars are likely to have a high $WTPS$ value. This is aligned with the majority of research that relies on star count alone as an indicator of popularity.

\begin{table}[ht]
\begin{center}
	\caption{Correlations of $WTPS$ and repository properties}
	\label{tab:measurements}
	\begin{tabular}{ll}
		\toprule
        Repository Metrics & $WTPS$     \\ \hline
        Age                   & 0.236089 \\
        Owner Followers       & 0.136390 \\
        Size                  & 0.000755\\
		\bottomrule
	\end{tabular}
\end{center}
\end{table}

\begin{table*}[ht]
\centering
\caption{The calculation of a regression line for popularity indicators forks, stars and watchers and $WTPS$ based on 4 different time intervals. }
\label{tab:differenttimes}
\begin{tabular}{l|cc|cc|cc|}
\cline{2-7}
\multicolumn{1}{l|}{}& \multicolumn{2}{c|}{Forks} & \multicolumn{2}{c|}{Stars} & \multicolumn{2}{c|}{Watchers} \\ \hline
\multicolumn{1}{|c|}{$WTPS$}&Slope&Coefficient&Slope&Coefficient&Slope& Coefficient\\ \hline
\multicolumn{1}{|l|}{t = 1 month} & 7.259     & 0.726          & 32.267     & 0.925         & 1.728      & 0.691            \\
\multicolumn{1}{|l|}{t = 3 weeks} & 10.595    & 0.727          & 47.311     & 0.929         & 2.518      & 0.688            \\
\multicolumn{1}{|l|}{t = 2 weeks} & 15.722    & 0.72          & 70.129     & 0.928         & 3.744      & 0.689            \\
\multicolumn{1}{|l|}{t = 1 week} & 31.142    & 0.726          & 138.961    & 0.928         & 7.411      & 0.689            \\ \hline
\end{tabular}
\end{table*}

 Table \ref{tab:measurements} shows the correlations; it is clear that there is not a strong relationship between age, size and owner's followers count and the popularity score $WTPS$. However, the age correlation to $WTPS$ is slightly higher than other metrics which may cause time to have some effect on the popularity score.

In order to find the best time interval length, we considered three more time intervals: 1 week, 2 weeks, and 3 weeks.  To do this, we computed the new $WTPS$  correlation to other popularity measurements using different time intervals; Table \ref{tab:differenttimes} presents these results. $WTPS$'s correlations to forks, stars and watchers are similar; shorter time interval lengths have a more positive slope. However, $WTPS$ is reasonably robust to interval changes.

\subsection{Popularity in Collaboration Networks}
Repositories can be ranked as being more or less popular when different indicators of popularity are considered.  Hence, if the GitHub repositories are mapped to a network, the network will exhibit different properties based on how popularity is scored.  To understand the effect of  different popularity indicators,  we consider different versions of the same network.

\begin{figure}[htbp]
	\begin{center}
		\includegraphics[width=0.3\textwidth]{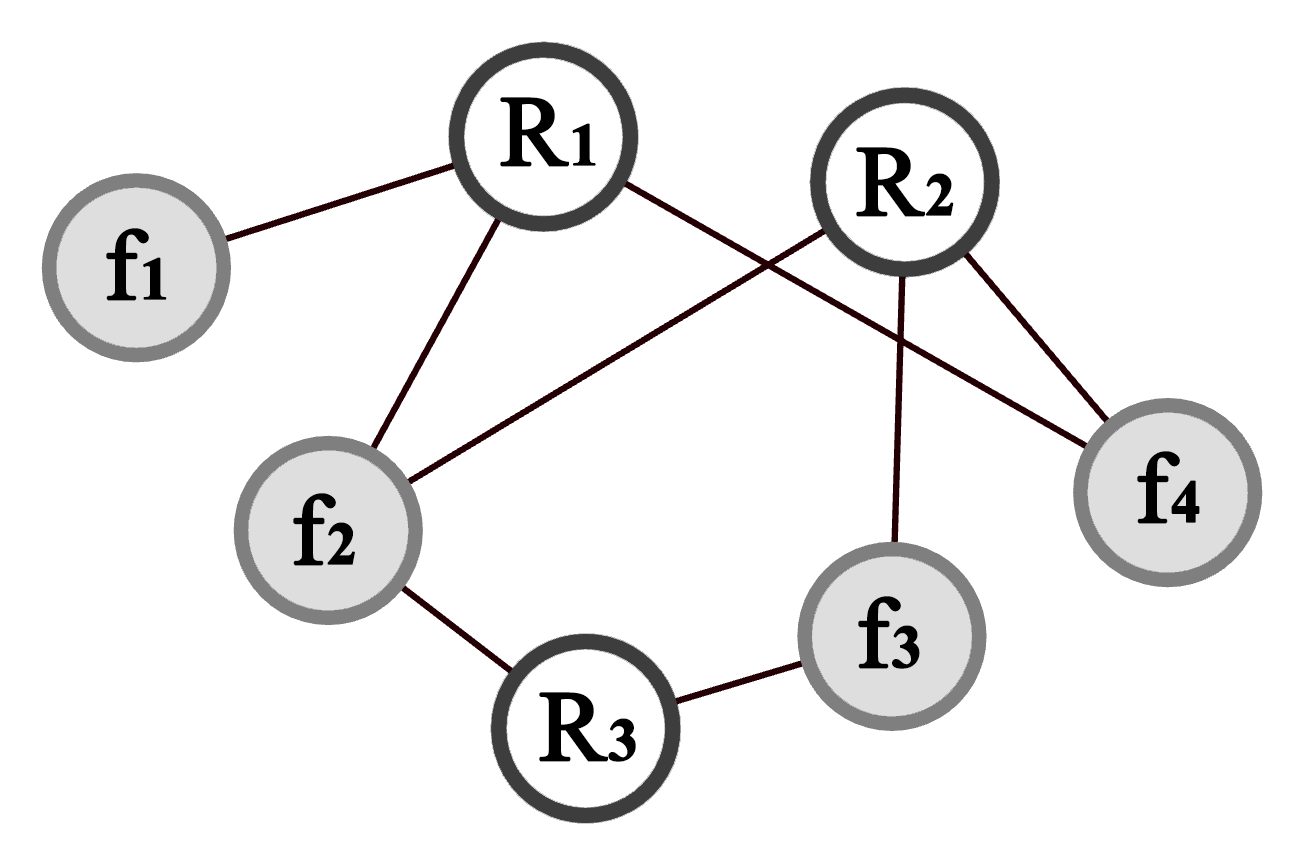}
	\caption{Subset of an example network: repository, owners' followers, and the repository-follower connections.}
	\label{networkSample}
	\end{center}
\end{figure}

\begin{figure}[htbp]
	\begin{center}
		\includegraphics[width=0.5\textwidth]{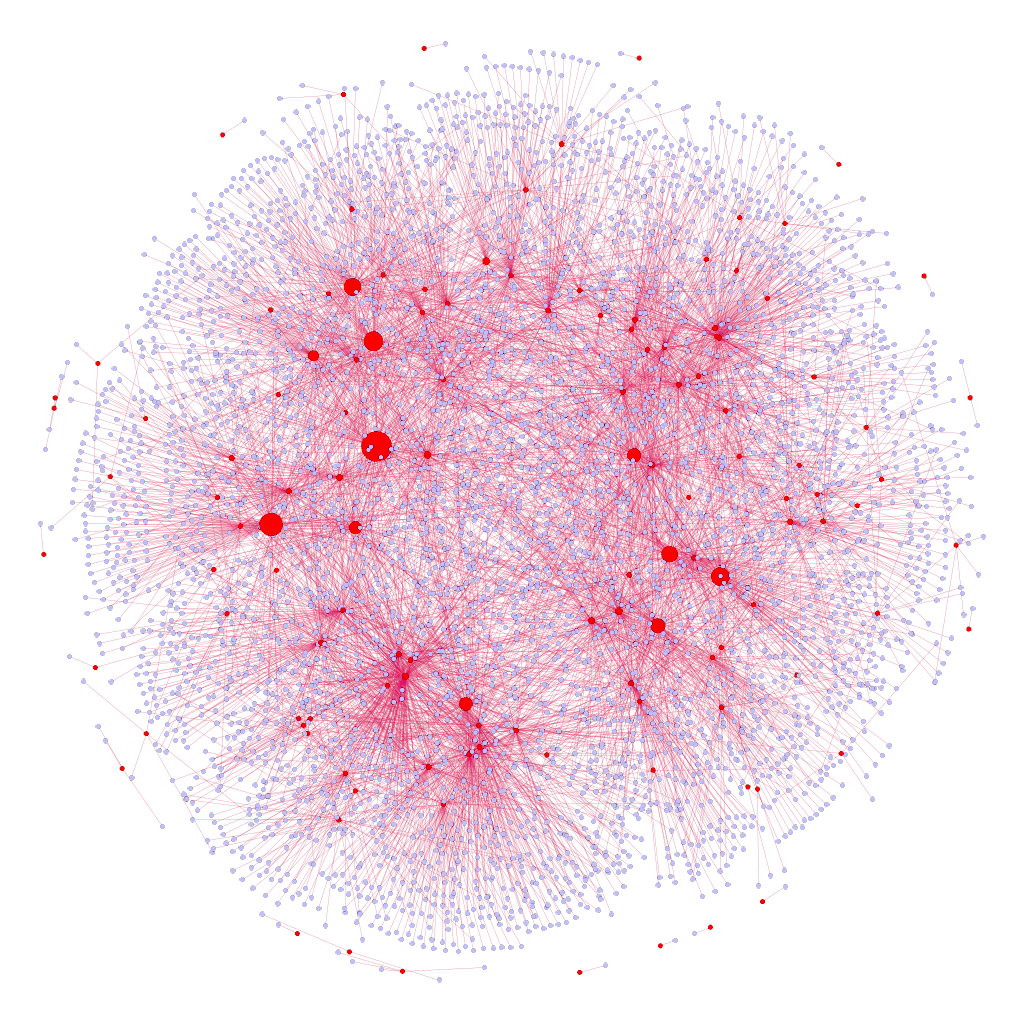}
	\caption{A sample graph of 5000 nodes and 5195 edges, where the red nodes represent the repositories and the blue nodes are the followers of the repository owners. The size of the repository nodes are proportional to $WTPS$ popularity.}
	\label{network}
	\end{center}
\end{figure}

The model that is defined here has two types of nodes: repositories, and the followers of repository owners where each repository is simply linked to its owners' followers. Figure \ref{networkSample} presents a sample subset of the model; as it is shown $R_1$ is linked to its owner's followers \{$f_1$, $f_2$, $f_4$\}, $R_2$ is linked to the followers \{ $f_2$,$f_3$,$f_4$ \} and $R_3$ is connected to \{$f_2$, $f_3$\}. Using this model, we construct a graph of repositories and their followers; Figure \ref{network} shows the graph where the large-sized nodes are repositories with higher $WTPS$.

In this section the aim is to determine whether selecting different popularity indicators affects the graph's structure. For this purpose we evaluate the impact of deleting nodes according to different popularity measures on the total clustering coefficient of the graph. Initially the total clustering coefficient of the graph is calculated, then on each step we remove a node from the graph that has the highest popularity.  At each step we select the node slated for removal based on different popularity measures: highest $WTPS$, highest stars, highest forks and highest watchers, and at each time step the clustering coefficient of the graph is recalculated.

\begin{figure}[htbp]
	\begin{center}
		\includegraphics[width=0.5\textwidth]{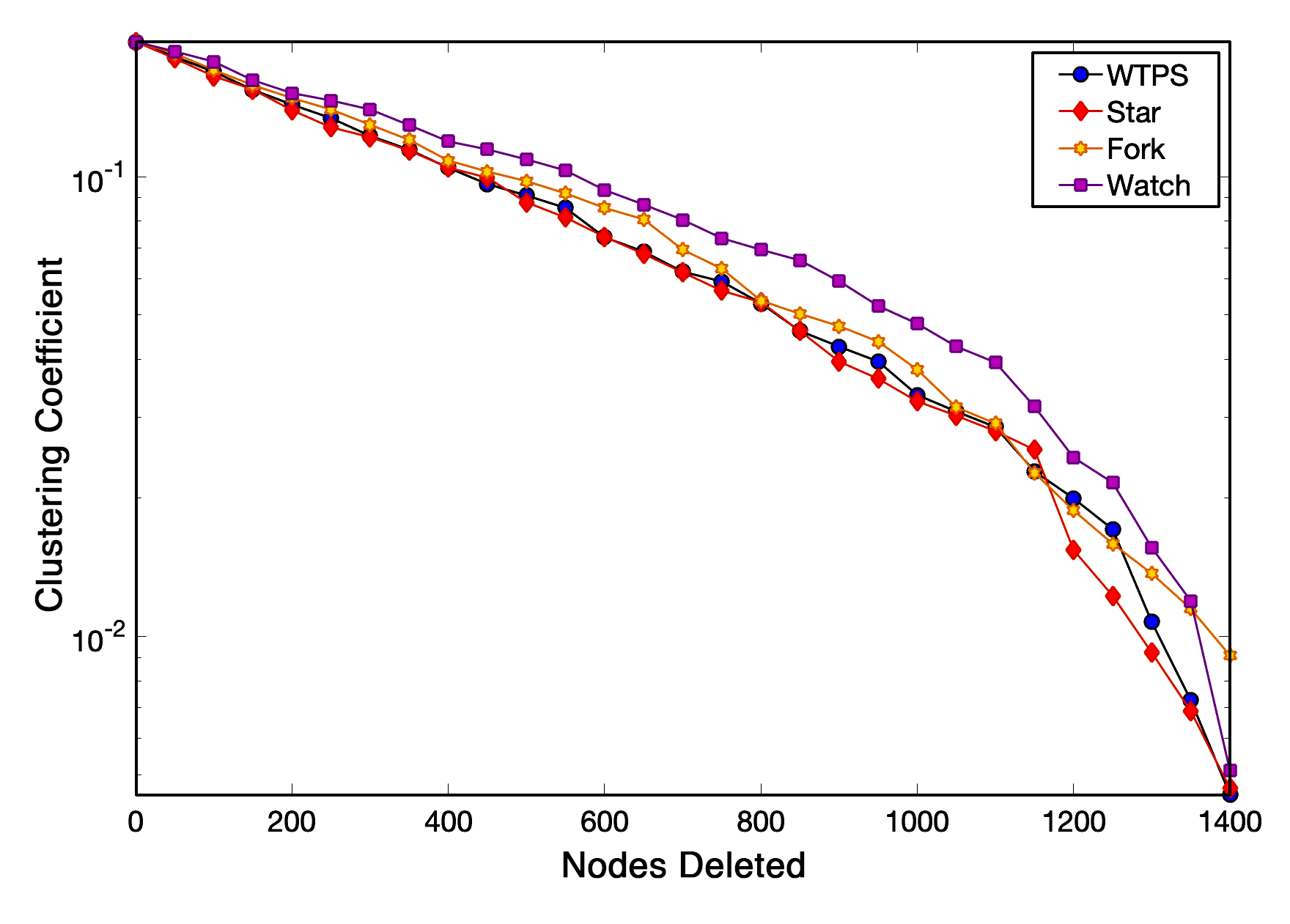}
	\caption{Repository deletion impact on one of the principal network characteristics: clustering coefficient.}
	\label{deletion_impact}
	\end{center}
\end{figure}

As shown in Figure \ref{deletion_impact} the clustering coefficient value becomes smaller after every deletion. This behavior is  similar for all four types of popularity-based node deletions. However, the trend is more similar when the node selection is based on either the highest $WTPS$ score or star count. Deletion based on the number of watchers leads to a slightly different result.  Thus we believe that research studies based on extracted GitHub networks should either use $WTPS$ or star counts when modeling information diffusion.

\section*{Conclusion}
This paper surveys many mechanisms for measuring the popularity of GitHub repositories.  Our proposed popularity score ($WTPS$) weights a repository's forks and stars, based on a comparison to the popularity gains made by other repositories during the same time period. The key advantage of $WTPS$ is that it is more robust to changes in social media platforms.  Since both the content and user population of social coding platforms increase over time, the best way to create a consistent measurement is to normalize increases vs. growth across the community.  Hence, our method requires gathering data from a large sample of GitHub repositories.  For this purpose the GitHub API was utilized to collect  collections of repositories.

 $WTPS$ was compared against two other popularity indicators, forks and stars, in order to examine their correlation. The results showed that $WTPS$ has a relatively high correlation coefficient with both indicators. However $WTPS$ tends to be more related to the value of the stars than forks.  In other words gaining more stars is more likely to increase $WTPS$, as compared to gaining forks.  Thus the usage of $WTPS$ is likely to produce results that are consistent with research studies that have relied on stars as a shorthand for popularity.

Furthermore we examined the correlation of $WTPS$ to several other repository measurements in order to find probable factors that may influence $WTPS$. To do so we performed a correlation analysis of $WTPS$ with age, total number of followers of a repository's owner and repository's size.  The results returned low correlation coefficients.  This is a useful property because it means that $WTPS$ can be used to fairly score a diverse group of repositories.  Among these three factors age had the highest correlation ($\sim$ +0.27). Even though the $WTPS$-Age correlation is not strong, it can mean that the older a repository is, the higher the likelihood of it having a greater $WTPS$.

We also defined a graph
that represents GitHub repositories and their owners' followers in which a follower of a repository owner is linked to the repository by an edge. The aim was to study the impact of removing popular nodes on the graph's total clustering coefficient.  Our popularity-based deletion process was conducted with $WTPS$, stars, forks, and watchers, in order to compare the impact on the graphs.  All four deletion processes led to close results. However the results of node deletions for $WTPS$ and stars were the most similar. 

To facilitate more research in this area, we have made our dataset publicly available at \url{https://drive.google.com/file/d/1T_lNPrIjmjnZr1ihw1hWan2N-mbNhwiR/view?usp=sharing}.

\section{Acknowledgments}
This work was partially supported by grant FA8650-18-C-7823 from the Defense Advanced Research Projects Agency (DARPA). The views and opinions contained in this article are the authors and should not be construed as official or as reflecting the views of UCF, DARPA, or the U.S. DoD.


%
%
\end{document}